\documentclass[a4paper,12pt]{amsart} 
\usepackage[utf8]{inputenc}
\usepackage{xcolor}
\usepackage{bm}

\RequirePackage{amsthm,amsmath,amsfonts,amssymb}
\RequirePackage[authoryear]{natbib}
\RequirePackage[colorlinks,citecolor=blue,urlcolor=blue]{hyperref}
\RequirePackage{graphicx}
\bibpunct{(}{)}{;}{a}{,}{,} 

\begin{document}

\title[A Bayesian spatio-temporal ECM]{A Bayesian spatio-temporal analysis of markets during the Finnish 1860s famine}

\author[T-M Pasanen]{Tiia-Maria Pasanen}
\address{Department of Mathematics and Statistics, University of Jyväskylä}
\email[Tiia-Maria Pasanen]{tiia-maria.h.pasanen@jyu.fi}

\author[M Voutilainen]{Miikka Voutilainen}
\address{Department of History and Ethnology, University of Jyväskylä}
\email[Miikka Voutilainen]{miikka.p.voutilainen@jyu.fi}

\author[J Helske]{Jouni Helske}
\address{Department of Mathematics and Statistics, University of Jyväskylä}
\email[Jouni Helske]{jouni.helske@jyu.fi}

\author[H Högmander]{Harri Högmander}
\address{Department of Mathematics and Statistics, University of Jyväskylä}
\email[Harri Högmander]{harri.hogmander@jyu.fi}

\maketitle 

\begin{abstract}

\noindent We develop a Bayesian spatio-temporal model to study pre-industrial grain market integration during the Finnish famine of the 1860s. Our model takes into account several problematic features often present when analysing multiple spatially interdependent time series. For example, compared with the error correction methodology commonly applied in econometrics, our approach allows simultaneous modelling of multiple interdependent time series avoiding cumbersome statistical testing needed to predetermine the market leader as a point of reference. Furthermore, introducing a flexible spatio-temporal structure enables analysing detailed regional and temporal dynamics of the market mechanisms. Applying the proposed method, we detected spatially asymmetric ``price ripples'' that spread out from the shock origin. We corroborated the existing literature on the speedier adjustment to emerging price differentials during the famine, but we observed this principally in urban markets. This hastened return to long-run equilibrium means faster and longer travel of price shocks, implying prolonged out-of-equilibrium dynamics, proliferated influence of market shocks, and, importantly, a wider spread of famine conditions.

\end{abstract}


\section{Introduction}

Analysing the co-movement of prices in multiple spatially connected regions is essential in gaining understanding of the behaviour, structure, and efficacy of the markets  \citep[e.g.,][]{Fackler2008, shin2010geospatial}. However, statistical modelling of such spatio-temporal phenomena can be challenging for various reasons.

Error correction models (ECM) \citep{engle1987co, alogoskoufis1991error} are a popular tool for studying the short-term co-movement and the long-run out-of-equilibrium dynamics in economics and other domains \citep[e.g.,][]{moller2014malthus, li2013towards, li2006tourism, nkang2006rice}, but traditional ECM assumes a single exogenous variable affecting another dependent variable, making it unsuitable for the context of joint modelling of spatio-temporal markets. While vector error correction models (VECM) \citep[see, e.g.,][]{juselius2006cointegrated} allow joint modelling of multiple interdependent time series, their use entails burdensome, error-prone, and potentially biasing testing for appropriate series in each co-integrated relationship, with additional difficulties due to potentially time-varying relationships \citep{giles1993, giles2012, gonzalo1998, hjalmarsson2010}.

In this paper, we study the integration of Finnish rye markets in the 1860s, during which Finland suffered the last peacetime famine in Western Europe. Same markets have been studied earlier in \citet{grada2001markets} using aggregated provincial data and a simple ECM approach where one province was fixed as a market leader and where the famine peak in 1867--1868 was accounted via a dummy variable. However, the use of aggregated data loses information about the smaller scale price variation and the concept of a market leader is somewhat artificial in poorly integrated markets which may contain multiple spatial equilibria \citep[e.g.,][]{studer2008india, chilosi2013europe, voutilainen2020multi}, especially during famines \citep[e.g.,][]{shin2010geospatial}, leaving market behaviour under such environments greatly understudied.  Instead, we construct a Bayesian spatio-temporal model which jointly models all regional price series without a need for a predetermined point of reference. Furthermore, as the famine was a protracted sosio-economic process punctuated by failing harvests throughout the 1860s, we use time-varying coefficients which require no pre-determined period for the famine and allow a smoothly shifting relationship between the spatial markets. While our model bears some similarity to SAR models \citep{ord1975estimation}, it provides a flexible and efficient estimation of the spatial correlation structures without a predefined spatial weight matrix. In addition, the model inherently separates the short-term and long-term regional relationships.

By allowing complicated feedback between the regions, we are able to study the overall stability of the market system. We observe that the out-of-equilibrium dynamics are contingent on the overall spatial organization of the markets. Speedier error correction facilitates transmission of price shocks, and regional feedback keeps the overall price system longer periods out of equilibrium. In essence, our findings cast doubt on whether speedier error correction between any pair of markets can solely be used to infer the behaviour of the market system as a whole due to the complex short-term and long-term interdependence within the markets on local and system-wide level. We also document the existence of ``price ripples'' \citep{seaman1980markets, devereux1988entitlements, hunter2019price} from a shock epicenter but observe a spatially asymmetric adjustment to price shocks. The majority of the increased spatial price transmission happened in urban--rural trade with some evidence that urban prices were increasingly influenced by the price level in the surrounding rural regions during the famine.

The paper is structured as follows. In Section \ref{sec:general_background}, we provide background for the Finnish famine of the 1860s and discuss the data used in this study. In Section \ref{sec:model}, we extend the traditional ECM to a spatio-temporal setting. In Section \ref{sec:results}, we implement the model to study the spatial price behaviour during the Finnish 1860s famine, leaving it for Section \ref{sec:discussion} to discuss the results and model extensions, and to conclude.

\section{Background and data}\label{sec:general_background}

The role of markets in safeguarding or depleting food security has remained in analytical focus, as recent famines \citep{maxwell2020determining} are characterised by a failure of markets to deliver access to food at affordable prices \citep{howe2004famine, devereux2009does, andree2020predicting}. While there is little disagreement on the theoretical benefits of a well-integrated market system (\citealp[e.g.,][]{van2007modelling, matz2015short, svanidze2019food, persson1999grain}, \citealp[see however, e.g.,][]{deng2007increased}), a distinction remains between the long-run gains and adverse behaviour of markets over shorter periods.

By common account, markets can influence famines in three ways: markets may 1) alleviate local food security problems through profit-seeking arbitrage, 2) balkanise and stop transmitting price information needed to signal outside traders of localised food shortages, and 3) make things worse by exporting food from a region of scarcity \citep[see, e.g.,][]{grada2005markets, grada2015eating, devereux2009does, devereux1988entitlements}. Most famines experienced during the past hundred years have been marked by disintegration and balkanization of market relationships \citep[e.g.,][]{shin2010geospatial, grada2015eating}, excessive price volatility \citep{araujo2016grain, ravallion1987markets, quddus2000speculative, devereux2007malawi, garenne2007atypical}, lack of accurate market information, and politically motivated hampering of market access \citep[e.g.,][]{olsson1993causes, waal1993war, macrae1992food, marcus2003famine, grada2015eating}.

This modern experience opposes the available evidence from the pre-industrial markets, which tended to function better during famines than in normal times \cite[e.g.,][]{grada2001markets, grada2005markets, grada2015eating, grada2002famine}. Several historical studies have concurrently suggested that the further back we go in time, the higher prominence low agricultural productivity and economic backwardness rise in explaining the occurrence and timing of famines \citep[e.g.,][]{campbell2010nature, hoyle2010famine, alfani2018timing, mishra2019drought}. With some notable exceptions \citep[e.g.,][]{slavin2014market}, these findings have poised skepticism that historical markets overwhelmed by supply shocks had the means or volumes needed to alleviate nationwide hardships. The Finnish 1860s famine has provided one of the original pieces of evidence that unlike their modern counterparts, the pre-industrial markets functioned well, even better during famines \citep{grada2001markets}. To contribute to the literature, we revisit this case of famine.

\subsection{The Finnish 1860s famine and data}\label{sec:background}

Held as the last peace-time population disaster of Western Europe, the Finnish 1860s famine occurred in an overwhelmingly agricultural economy in the hinterland of an industrializing continent and claimed close to $200{,}000$ lives (in excess mortality and in absence of births) out of a population of $1.8$ million. With the proximate cause in a substantial drop in harvests (on aggregate losses over $50$\% depending on location and grain variety) which could not be supplemented with imports, the famine mortality was greatly amplified by the pre-famine increase in vulnerable population stemming from high population growth, concentrated land ownership, and from lack of urban and industrial employment opportunities \citep{voutilainen2016poverty}.

In the mid-19th century, although Finland was sparsely settled and often self-reliant regarding grain output, food markets were sufficiently developed. This was true particularly in the historically well-off areas of Southwest Finland as well as in the southeastern parts of the country that laid close to the grain markets of St. Petersburg. Grain sales were an important source of income especially in the South-West Finland, and a substantial amount of grain consumed during harvest failures was purchased beyond local markets all over the country \citep{soininen1974vanha}. Due to historical trade connections between rural areas and towns and due to geographical constraints on movement (especially water ways), the regional nature of the inland trade remained in place well into the mid-19th century. Towns and town merchants were responsible for the largest trade volumes, and towns due to their livelihood structure had the most developed markets for food. Rural trade was liberalised in 1859, but it was mostly conducted to meet households' day-to-day needs and not used for large scale rural food retailing \citep{voutilainen2020multi}.

Our analysis of the food markets is based on the price of rye---the most widely traded and most voluminously produced grain of the time \citep{soininen1974vanha}. The price data were obtained from local reports sent to the Department of Finance and stored in the Finnish National Archives in Helsinki.\footnote{We are grateful to Kari Pitkänen for allowing us to use the data. Additional archival work to fill in the gaps in the data was conducted for the purpose of this study.} Starting in 1857, the officials of $48$ administrational districts and in $32$ town administrations were required to report monthly price denominations of the local prices of various commodities. The prices were market prices and best ones available from the era \citep{pitkanen1995deprivation}. The data have been previously analysed by \citet{grada2001markets}, who---in contrast to our approach---divided the data in eight provincial aggregates downgrading regional variance and leaving, among other things, the within-province heterogeneity and urban--rural division unhandled. The higher spatial resolution of this study is more in resonance with livelihood and market regions \citep{voutilainen2020multi, maxwell2020determining}, thereby providing more accurate measurement of the actual prices faced by the contemporary people.

\begin{figure}[ht!]
    \centering
    \includegraphics{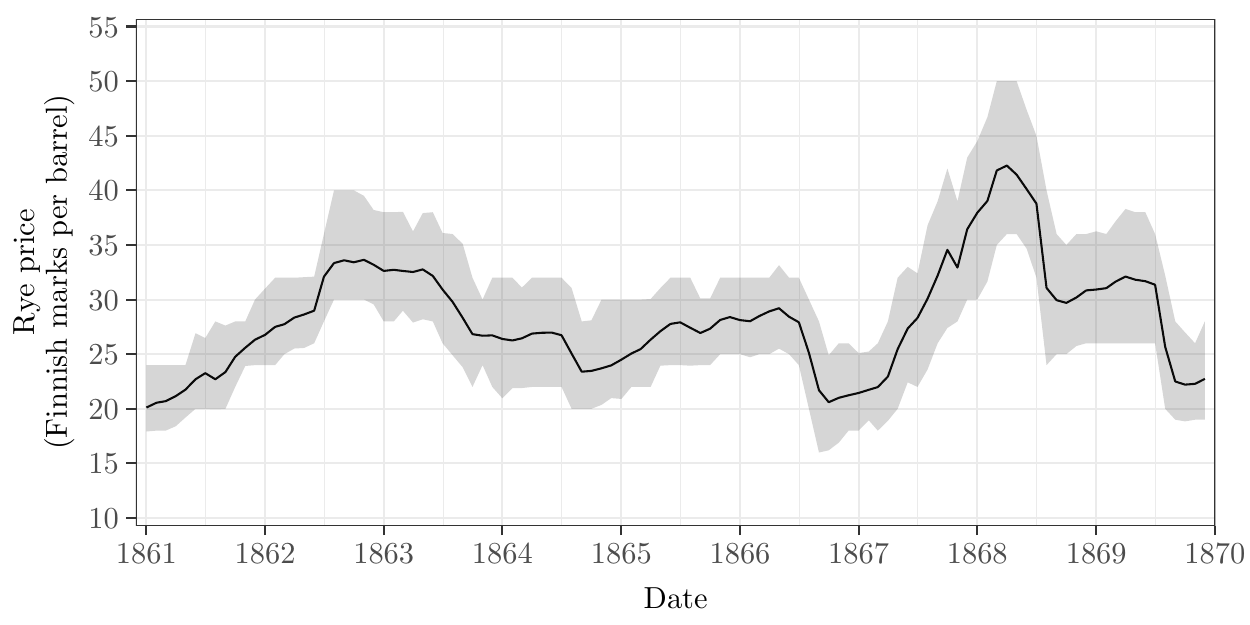}
    \caption{Rye price in Finland during 1860s as Finnish marks (FIM) per barrel. Average as black curve and $95$\% quantile interval as shaded area.}
    \label{fig:price}
\end{figure}

\autoref{fig:price} plots the aggregate price movement between January 1861 and December 1869, which is the time span of the data we analyse. In the beginning of our time span, in 1861, the coefficient of variation of regional rye prices was about $0.1$, roughly at par with the contemporary Western European economics, and signaling reasonably well integrated markets in terms of sigma convergence \citep[e.g.,][]{federico2021european}. The 1860s famine, which peaked in terms of mortality in the winter and spring of 1867--68, was foreshadowed by multiple harvest failures earlier in the decade. Harvest failures of 1862, 1865, and especially the famine escalating one in 1867, were tailed by a substantial increase in grain prices over the following winter.

\section{Spatio-temporal error correction model}\label{sec:model}

If markets are perfectly integrated, the prices in any two places should follow the so-called law of one price. Under these circumstances, a price differential will signal opportunities for arbitrage up to the point when prices are either identical or differ to reflect the transportation costs \citep{studer2008india, grada2015eating}. One of the implications of this is that the prices in two markets are characterised by a long-run relationship, deviations from which vanish by ensuing arbitrage \citep{grada2001markets, studer2008india}. This is customarily studied using an error correction approach \citep[e.g.,][]{alogoskoufis1991error, kennedy2008guide}. Considering two (log-)price time series $\bm{y} = (y_1, y_2, \dots, y_T)$ and $\bm{x} = (x_1, x_2, \dots, x_T)$, where $T$ is the number of time points, an ECM is traditionally represented as
\begin{equation}\label{eq:ecm}
  y_t - y_{t-1} = \alpha + \beta (x_t - x_{t-1}) + \gamma (y_{t - 1} - x_{t - 1}) + \epsilon_t,
\end{equation}
where $\alpha$ stands for a common trend, $\beta$ denotes the short-run effect of changes in the reference series $x$ on $y$, and $\gamma$ denotes the effect of the difference between $y$ and $x$ and $\epsilon_t \sim \text{N}(0, \sigma_{y}^2)$.

In normally functioning markets, we expect that $0 < \beta < 1$ and  $-1 < \gamma < 0$. The short-term coefficient $\beta$ captures the price co-movement between the two markets. The error correction term $\gamma$ is the long-term adjustment parameter: the share of disequilibrium (“error”) eradicated in each period. The higher the efficiency of the markets---that is, the closer $\gamma$ is to $-1$---the larger the proportion of the error vanishing between time points $t$ and $t+1$, and therefore the quicker the emerged disequilibrium is arbitraged away.

To provide a flexible method for multiple spatially related series, we generalise the common error correction equation \eqref{eq:ecm} in the following way. Consider a geographic region that can be partitioned into sub-regions denoted by $i = 1, 2, \dots, N$. In our case, we regard regions sharing borders as neighbours, even though some other condition for neighbourhood could be used as well. We denote the neighbourhood of a site $i$ with $J_i$, and an individual neighbour with $j \in J_i$. There is a location specific price $\mu_{i,t}$ of the product for each region and each time point. The prices for all the regions at one time point are denoted by $\bm{\mu}_{t} = (\mu_{1, t}, \mu_{2, t}, \dots, \mu_{N, t})^T$.

Now, if we denote the $y_t$ and $x_t$ in the original ECM by $\mu_{i,t}$ and $\mu_{j,t}$, respectively, the model in the case of only one neighbour looks like
\begin{equation*}\label{eq:one_neigh}
  \mu_{i, t} = \mu_{i, t-1} + \alpha + \beta_{i, j} (\mu_{j, t} - \mu_{j, t-1}) + \gamma_{i, j} (\mu_{i, t-1} - \mu_{j, t-1}) + \epsilon_{i, t}.
\end{equation*}
The errors $\epsilon_{i,t}$ are assumed to be Gaussian. This model, in turn, can easily be generalised for an arbitrary number of neighbours by adding new terms to all the neighbouring sites in question; thus, we obtain
\begin{equation*}\label{eq:many_neigh}
  \mu_{i, t} = \mu_{i, t-1} + \alpha + \sum_{j \in J_i} \beta_{i, j} \left(\mu_{j, t} - \mu_{j, t-1} \right) + \sum_{j \in J_i} \gamma_{i, j} \left(\mu_{i, t-1} - \mu_{j, t-1} \right) + \epsilon_{i, t}. \\
\end{equation*}

We now consider all the regions and their possible connections simultaneously by combining all the site specific models and including the spatial dependency structure in its entirety into a single model frame. The equation above can be written for all regions simultaneously with a matrix notation as
\begin{equation*}\label{e:many_matrix}
   \bm{\mu}_{t} = \bm{1}\alpha + B \bm{\mu}_{t} + D \bm{\mu}_{t-1} + \bm{\epsilon}_{t}.
\end{equation*}
The term $\bm{1}$ denotes an all-ones vector of length $N$, and the entries of the $N \times N$ matrix $B$ are
\begin{equation*}
    b_{i,j} = 
    \begin{cases}
            \beta_{i,j}, &\text{if $i$ and $j$ are neighbours, i.e., } j \in J_i \\
            0,           &\text{otherwise.}
    \end{cases}
\end{equation*}
The $N \times N$ matrix $D$, in turn, has the entries
\begin{equation*}
    d_{i,j} = 
    \begin{cases}
        -(\beta_{i,j} + \gamma_{i,j}), &\text{if $i$ and $j$ are neighbours, i.e., } j \in J_i \\
        1 + \sum\limits_{j' \in J_i} \gamma_{i,j'}, &\text{if $i$ = $j$}\\
         0, &\text{otherwise.}
    \end{cases}
\end{equation*}
With a short notation $$ \bm{\eta}_t = \bm{1}\alpha + D \bm{\mu}_{t - 1} $$ the previous equation can be expressed as
\begin{equation*}\label{eq:model}
  \bm{\mu}_{t} \sim \text{N}((I - B)^{-1} \bm{\eta}_t, \sigma_{\mu}^2 (I - B)^{-1} ((I - B)^{-1})^T). \\
\end{equation*}
Instead of explicitly specifying the cointegration relationships, reference series, and other complex interdependency structures between the regions, we started with a generalization of a simple ECM with a fixed neighbourhood structure and, perhaps surprisingly, arrived at a model that bears similarity with simultaneous autoregressive models (SAR) \citep{ord1975estimation, anselin1988spatial}, which are widely used in spatial statistics. Compared to a typical SAR model in which the neighbourhood is defined via $B = \rho W$ with a spatial dependency parameter $\rho$ and a fixed spatial weight matrix $W$, here the non-zero elements of $B$ are unknown parameters with an interpretation corresponding to the original ECM formulation. The exogenous variables commonly used as predictors in a SAR model are replaced here with a lagged dependent variable $\mu_{t-1}$, which follows directly from the above expansion of the original ECM. The advantage of our formulation is that the underlying cointegration relationships are now implicitly included by the spatial spillover of the SAR structure. This enables a situation where a region $j$ at time $t$ may have an effect on a region $i$ at the next point of time $t+1$, even though the regions $i$ and $j$ are not direct neighbours. An extension of standard SAR model with heterogenous spatial lag coefficients was developed in \citep{Aquaro2020}, but this model does not decompose the spatial and temporal relationships into short- and long-term effects as our ECM-inspired formulation, and thus lacks some of the interpretative power of our model.

Whether the error correction parameters change during famines has been studied using a predetermined famine period demarked by a dummy variable  \citep[e.g.,][]{grada2001markets, grada2005markets}. This, however, risks disregarding the regionally desynchronised evolution of famine conditions and enforcing a ``cliff-edge'' of famine periodization \citep[e.g.,][]{maxwell2020determining}. Ideally, the model should provide estimates for time-dependency independent of pre-assumptions. Therefore, we allow the parameters $\alpha$ and $\gamma$ to vary in time. Previous literature has introduced time-dependency to the short-run coefficients $\beta$ \citep[e.g.,][]{ramajo2001time, li2006tourism}; however, regarding market integration, the interest lies in the market adjustment parameter $\gamma$. Furthermore, we introduce time-dependency to the parameter $\alpha$ to account for common nationwide shocks that may surface as an increased price co-movement during the famine \citep[e.g.,][]{grada2001markets}, alongside a common (irregular) seasonal variation due to harvests. Since we do not wish to assume all the sites are identical with respect to $\alpha$, we introduce also a site specific dependency of the nationwide trend, leading to
\begin{equation}\label{eq:latent}
  \bm{\mu}_{t} = \bm{\lambda}\alpha_{t-1} + B \bm{\mu}_{t} + D_{t-1} \bm{\mu}_{t-1} + \bm{\epsilon}_{t}.
\end{equation}
The term $\bm{\lambda}$ denotes a vector of the site specific coefficients $\lambda_i$ that depict how strongly the site $i$ depends on the common $\alpha_{t-1}$. $B$ is as before. The matrix $D_{t-1}$ depends on time through $\gamma_{t-1}$, meaning that instead of having only one matrix compressing the information about the whole period, there is an individual coefficient matrix for each time point. We model the time-varying parameters $\gamma_{i,j,t}$ as random walks and coefficients $\alpha_{t-1}$ as a stationary first-order autoregressive (AR($1$)) process.

In practice, our data have $T = 108$ time points and $N = 80$ regions with $298$ neighbour pairs. It is plausible that our historical price data contains inaccuracies, for example, due to the temporal coarseness of the measurements or errors in the reporting process itself. The latter is underlined since our data are not complete, but $5$\% of the observations are missing, probably attributed to the absence of actual price reports or the loss of these reports in the archiving process. Therefore, it is natural to think that the actual price development is a latent process from which we have only noisy observations. The observed log-prices from $N$ sites at the time point $t = 1, 2, \dots, T$ form a vector $ \bm{y}_{t} = (y_{1, t}, y_{2, t}, \dots, y_{N, t})^T $ for each time point. They are considered to be realizations from a Gaussian distribution
\begin{equation}\label{eq:obs}
  \bm{y}_{t} \sim \text{N}(\bm{\mu}_{t}, \sigma_{y}^2 I), 
\end{equation}
with the expected value $\bm{\mu}_{t}$ as in \eqref{eq:latent}. The variance $\sigma_y^2$ is assumed to be constant for all sites and times, and $I$ denotes an identity matrix. The variables $\bm{\mu}_t$ are the true unobserved prices on the latent level and $\bm{y}_t$ are our observed data.

\section{Results}\label{sec:results}

We estimate the model based on Equations \eqref{eq:latent} and \eqref{eq:obs} using a Bayesian approach.\footnote{The material to reproduce the analysis is available on \url{https://github.com/tihepasa/bayesecm}.} This allows the estimation of complex hierarchical models with prior information and missing data, especially as we wish to take into account the uncertainty due to parameter estimation while interpreting the model parameters (or their arbitrary functions, as in Section \ref{sec:maps}). Bayesian approach also naturally accommodates the possibility that the time-varying components of our model ($\gamma$, $\alpha$) are (nearly) constant in time by averaging over the uncertainty of the respective standard deviation parameters (see below). Following theoretical assumptions underlying price transmission ECMs, we restrict the coefficients $\beta_{i,j}$ to be positive and $-1\leq \gamma_{i,j,t} \leq 0$ for all $i,j$ and $t$, and let the unconstrained $\tilde{\gamma}_{i,j,t} = \operatorname{logit}(-\gamma_{i,j,t})$ follow linear Gaussian random walks (with respect to $t$). 
As a prior for region-specific coefficients $\lambda_i$, we define the marginal distribution of each $\lambda_i$ as N$(1, \sigma_\lambda^2)$, with an additional constraint that the mean of coefficients is exactly 1. This ensures that the product $\bm{\lambda}\alpha_{t-1}$ is identifiable \citep{bai2015} while keeping $\bm{\alpha}$ interpretable. 

The prior distributions of our full model are
\begin{align*}
  \alpha_1 & \sim \text{N}\left(\frac{c_{\alpha}}{1 - \phi}, \left(\frac{\sigma_{\alpha}}{\sqrt{1 - \phi^2}}\right)^2\right), \\
  \alpha_t & \sim \text{N}(c_{\alpha} + \phi \alpha_{t-1}, \sigma_{\alpha}^2), \ \text{for } t > 1, \\
  c_{\alpha} & \sim \text{N}(0, 0.1^2), \\
  \phi & \sim \text{Beta}(2, 2), \\
  \sigma_{\alpha} & \sim \text{Gamma}(2, 100), \\
  \lambda_i & \sim \text{N}(1, \sigma_{\lambda}^2), \ \text{for all } i, \text{ given } \frac{1}{N}\sum_{i=1}^N \lambda_i = 1,\\
  \sigma_{\lambda} & \sim \text{N}(0, 0.5^2)[0,],\\
  \beta_{i, j} & \sim \text{Gamma}(0.5, 2) \ \text{for all } i, j, \\
  \tilde{\gamma}_{i, j, 1} & \sim \text{N}(-2, 2^2) \ \text{for all } i, j, \\
  \tilde{\gamma}_{i, j, t} & \sim \text{N}(\tilde{\gamma}_{i, j, t-1}, \sigma_{\gamma}^2) \ \text{for all } i, j, \text{and} \ t > 1, \\  
  \sigma_{\gamma} & \sim \text{Gamma}(2, 10), \\
  \mu_{i, 1} & \sim \text{N}(3, 0.5^2) \ \text{for all } i, \\
  \sigma_{\mu} & \sim \text{Gamma}(2, 20) \text{, and}\\
   \sigma_y & \sim \text{Gamma}(2, 20),
\end{align*}
where N$(\cdot, \cdot)[0,]$ denotes truncated (at zero) normal distribution.

The priors were chosen based on the approximate prior scale of the variables and then adjusted based on the initial Markov chain Monte Carlo (MCMC) runs for enhanced computational efficiency of the final MCMC sampling. However, the chosen priors did not have strong influence on the posterior estimates compared to more diffuse choices (see supplementary material on GitHub), so they can be regarded as only weakly informative.

The model was estimated with \textit{rstan} \citep{rstan}, which is an R interface \citep{r} for the probabilistic programming language Stan for statistical inference \citep{stan}. 
The samples were drawn using the NUTS sampler \citep{nuts} with four chains, each consisting of $8{,}000$ iterations, with the $3{,}000$ first discarded as a warm-up. The total computation time with parallel chains was about thirteen hours. The effective sample sizes were approximately between $700$ and $45{,}000$, with the most inefficient estimation on the strongly correlating parameters $\sigma_y$ and $\sigma_{\mu}$.

During the estimation of the model, all regions were treated equally without defining whether they were rural districts or urban towns, as we did not have clear a priori knowledge about the potential differences or similarities in say $\beta$ or $\gamma$ coefficients between or within these groups. However, there was noticeable posterior evidence that the rural and urban regions behave somewhat differently, and hence we present the results taking this grouping into account. 

Instead of describing all region-specific results, in some instances we focus on three example regions, namely the rural districts of Ilmajoki and Masku, and the town of Viipuri. These choices were based on the estimated model; these regions were found to be important drivers of the prices in their neighbours. Ilmajoki is located on the western coast, whereas Masku lies within South-West Finland, where the vast share of marketed grain surplus was produced \citep{soininen1974vanha}. In addition, the choice of Viipuri was motivated by initial observations by \citet{grada2001markets}, who stressed the importance of the wider southeastern Viipuri province as a market leader, possibly due to its proximity to the Russian grain markets.

\begin{figure}[h!]
    \centering
    \includegraphics{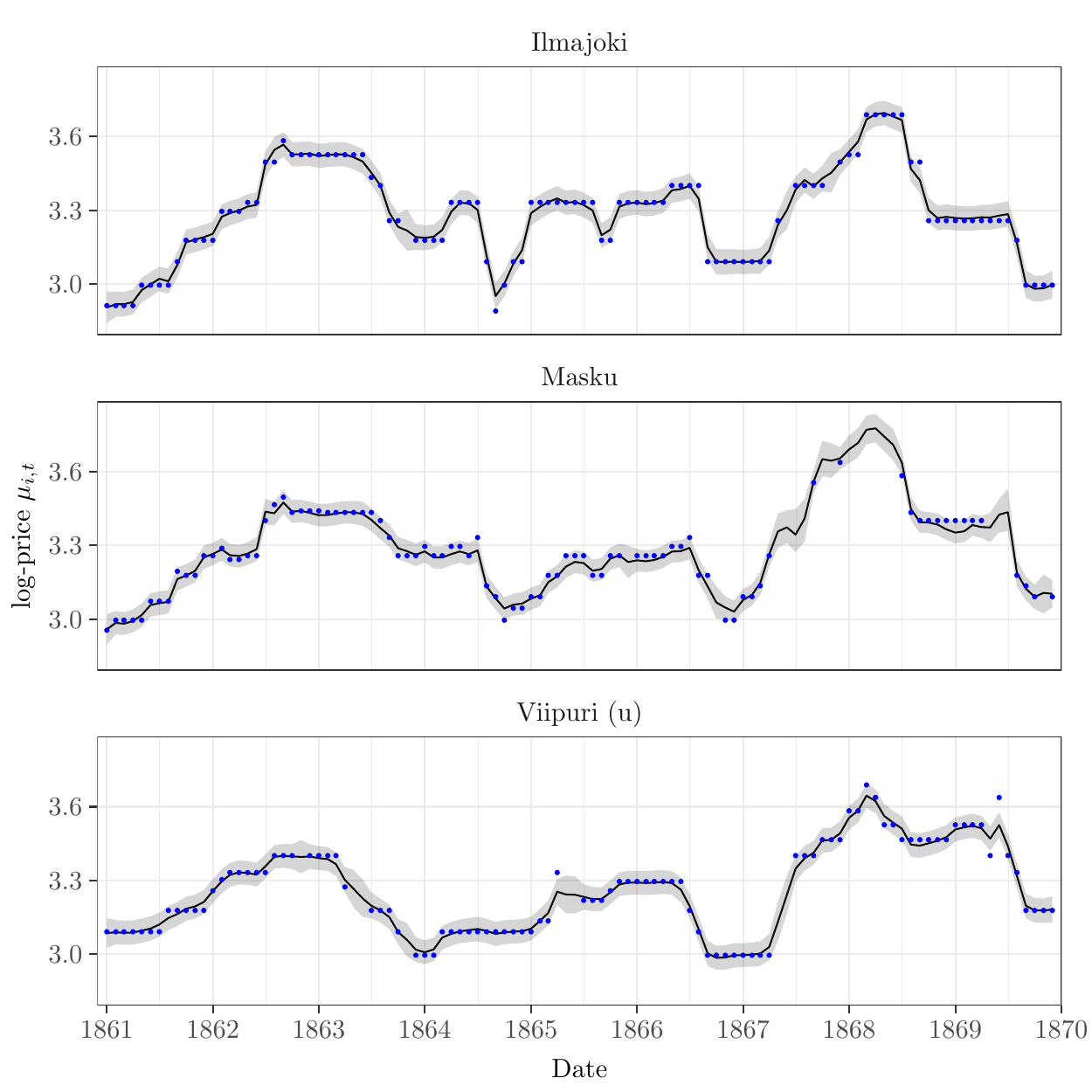}
    \caption{Posterior mean and $95$\% posterior interval for the unobserved log-price $\mu_{i,t}$ in Ilmajoki, Masku, and Viipuri (urban) over the time period under study. The dots represent the observed log-prices $y_{i,t}$.}
    \label{fig:mu}
\end{figure}

\autoref{fig:mu} shows  examples of estimates for the unobserved log-prices $\mu_{i,t}$ and corresponding $95$\% posterior intervals for Ilmajoki, Masku, and Viipuri. In each instance, the estimates smoothly follow the observed prices. Reasonable estimates were also obtained for missing observations. The estimate of the parameter $\sigma_{\mu}$, related to the deviation of the unobserved log-prices, is $0.038$ with a $95$\% posterior interval $[0.035, 0.041]$. The standard deviation of the observed log-prices, $\sigma_y$, is $0.036$ with a $95$\% posterior interval $[0.033, 0.038]$.

Perhaps a natural alternative to our analysis would be to model latent log-prices via a SAR model.
To this end, we defined $\mu_t - \mu_{t-1} = \rho W (\mu_t - \mu_{t-1}) + \epsilon$, leading to $\mu_t \sim N(\mu_{t-1},\sigma_{\mu}^2(I -\rho W)^{-1}((I -\rho W)^{-1})^T)$,
where $W$ is the fixed adjacency matrix after spectral normalization \citep{kelejian2010specification}, and $-1 < \rho < 1$ is the unknown spatial dependency parameter. Thus, in addition to latent $\mu$, this model has only three unknown parameters, $\sigma_y$, $\sigma_{\mu}$, and $\rho$. These were estimated as 0.034 ([0.032, 0.035]), 0.055 ([0.054, 0.057]) and 0.778 ([0.755, 0.801]), respectively, implying strong spatial dependence and larger unexplained variation in the latent log-prices than our main model. The estimates of the $\mu_{i,t}$ resembled those in \autoref{fig:mu} except that the posterior intervals were wider compared to our main model (see supplementary material on GitHub). Overall, such a simplification does not provide the necessary information to identify the spatio-temporal features we are interested in, e.g., the short and long-term decomposition. Thus, we do not consider it more thoroughly here. Note, however, that this alternative is a submodel of our more detailed approach.

\subsection{Coefficient estimates}

\begin{figure}[h!]
    \centering
    \includegraphics{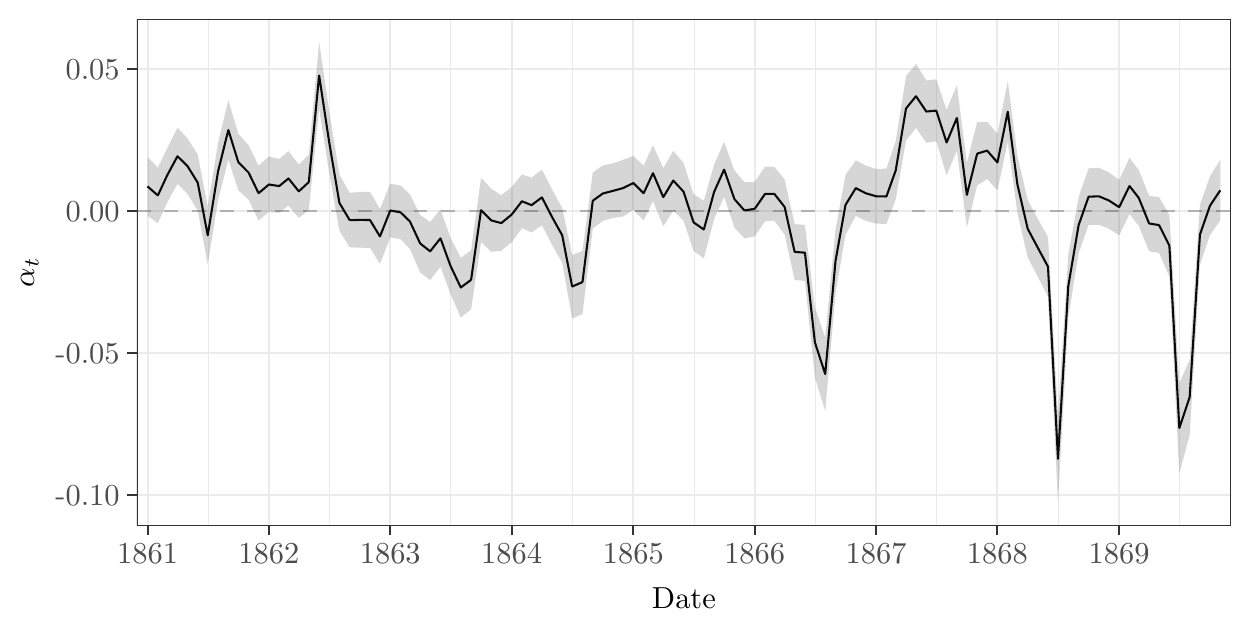}
    \caption{Posterior mean and $95$\% posterior interval for the trend $\alpha_t$ over the period under study.}
    \label{fig:alpha}
\end{figure}

\autoref{fig:alpha} shows the common time-varying  component $\alpha_t$, which  captures not only the inherent seasonal variation but also other unidentified nationwide price variation, for example due to poor harvests. The parameters related to the AR process of $\alpha$ are estimated as $c_{\alpha} = 0$ with a $95$\% posterior interval $[-0.003, 0.004]$, $\phi = 0.619$ ($[0.461, 0.766]$), and $\sigma_{\alpha} = 0.017$ ($[0.013, 0.021]$). In general, the $\alpha_t$ varies around zero, though exhibiting some larger swings. These are mainly price increases due to harvest failures (e.g., in autumn 1862 and 1867) and price drops due to relatively successful new harvests (e.g., in autumn 1868). \autoref{fig:alpha} also shows the general tendency of prices to rise between two harvests (see also \autoref{fig:price}). During the famine pinnacle of 1867--1868 this tendency was particularly pronounced.

\begin{figure}[h!]
    \centering
    \includegraphics{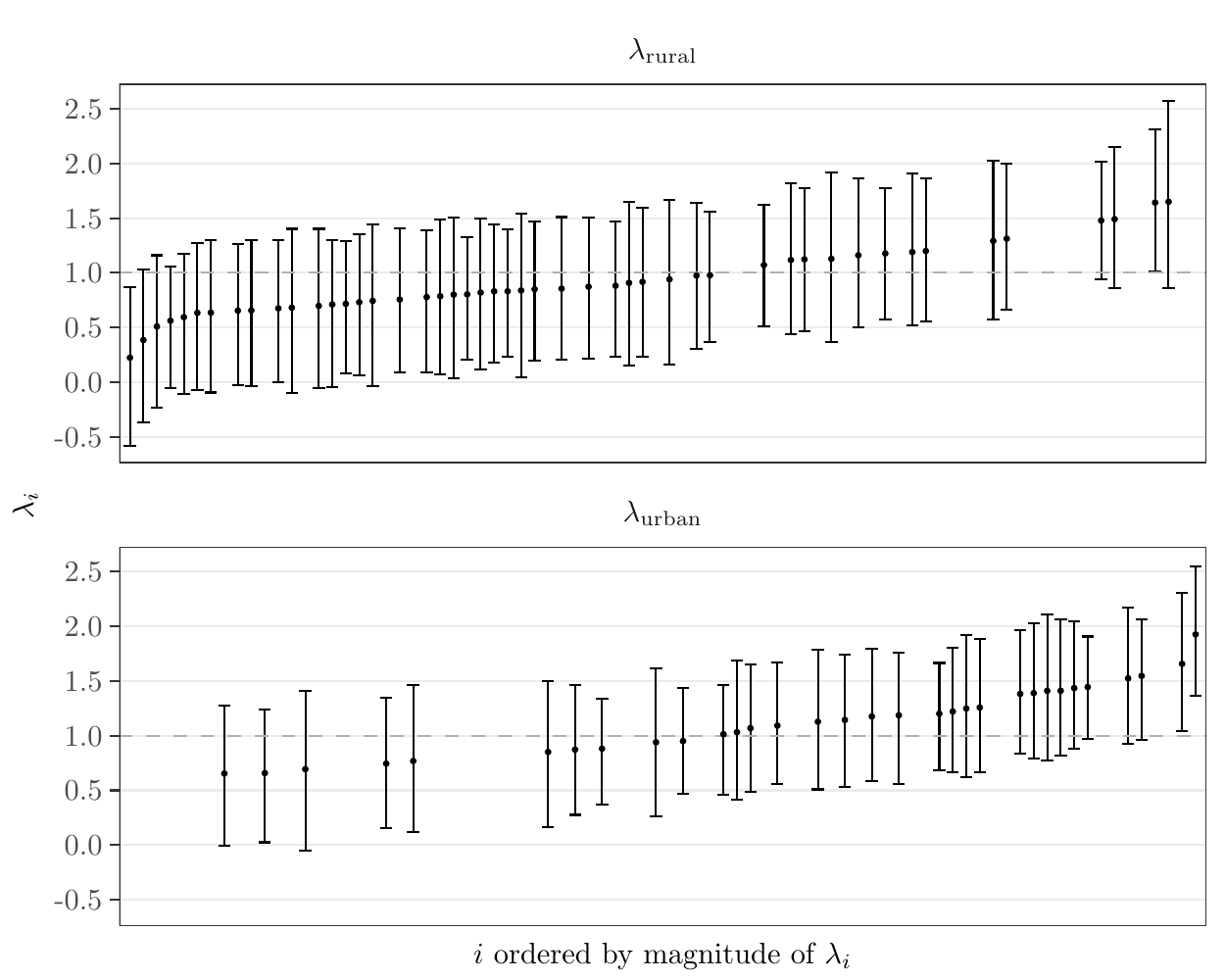}
    \caption{Posterior median and $95$\% posterior interval for the site specific coefficients $\lambda_i$ by urban and rural groups. The estimates are ordered by their size in order to illustrate the difference between coefficients of urban and rural districts.}
    \label{fig:lambda}
\end{figure}

The site specific coefficients $\lambda_i$ allowing spatial level differences from the nationwide trend $\alpha_t$ are plotted in \autoref{fig:lambda}. The posterior medians vary between $0.224$ and $1.650$ among the rural sites (upper panel) and $0.654$ and $1.927$ among the urban sites (lower panel). The posterior mean of the standard deviation $\sigma_{\lambda}$ is $0.463$ with a $95$\% posterior interval $[0.316, 0.632]$. Overall, the coefficients corresponding to the urban towns are more concentrated to the high end of the range of $\lambda$s, suggesting that urban towns reacted more strongly to the nationwide shocks captured by $\alpha$. This stronger dependency could be related to the fact that towns had to purchase grain consumed from surrounding rural regions.

\begin{figure}[h!]
    \centering
    \includegraphics{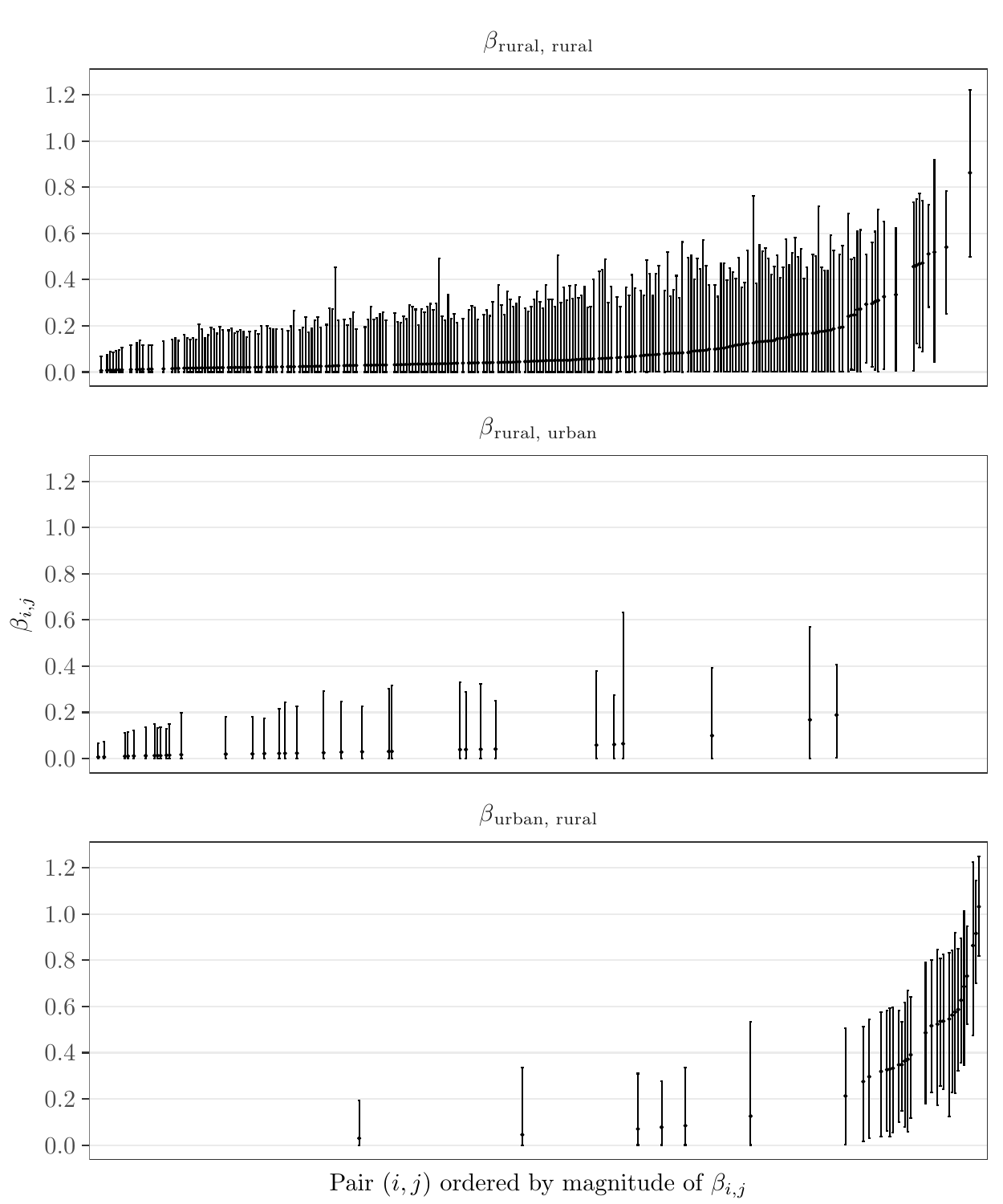}
    \caption{Medians and $95$\% posterior intervals of each coefficient $\beta_{i,j}$ grouped by top: rural districts with rural neighbours, middle: rural districts with urban neighbours, and bottom: urban districts with rural neighbours.}
    \label{fig:beta}
\end{figure}

\autoref{fig:beta} illustrates the short-term coefficients $\beta_{i,j}$ for each of the $298$ neighbour pairs. We group also the coefficients $\beta_{i,j}$: rural regions and their rural neighbours, urban locations and their surrounding rural regions. The upper panel reports all rural--rural pairs (both $\beta_{i,j}$ and $\beta_{j, i}$ included), the middle panel reports the rural--urban pairs (i.e., $\beta_{\text{rural, urban}}$), and the lower panel reports the urban--rural pairs ($\beta_{\text{urban, rural}}$). The dots represent the posterior medians of the individual coefficients $\beta_{i,j}$, which are all between $0.006$ and $1.020$. 

As is visible from \autoref{fig:beta}, the short-run coefficients $\beta_{i,j}$ are of similar magnitude both in rural--rural and in rural--urban pairs (upper and middle panel). The vast majority of these coefficients are small and fall between $0$ and $0.2$, the latter implying a price increase of $0.2$\% in location $i$ in response to a $1$\% increase in location $j$. This suggests that the short-run co-movement of prices (beyond aggregate fluctuations captured in $\alpha_t$) in rural--rural and rural--urban pairs were weak.

Interestingly, we detected asymmetry in the price co-movement. The urban--rural pairs (lower panel) show that urban prices were generally more sensitive to follow the price development of the surrounding rural region than vice versa. Majority of the coefficients $\beta_{\text{urban, rural}}$ lies between $0.3$ and $0.6$, with multiple urban locations with coefficients $\beta_{i,j}$ above $0.5$. This implies a larger than $0.5$\% increase in prices in response to $1$\% increase in the price level in the surrounding rural area.

This means that urban people were susceptible to market-transmitted shocks; conversely, rural prices were merely marginally affected by the urban demand pressure. In all likelihood, this is because urban consumers more frequently resorted to market purchases to obtain the grain consumed, thereby inducing a more developed market system in towns than in the rural regions  \citep{devereux1988entitlements}.  

The long-term market adjustment is captured by the parameters $\gamma_{i,j,t}$. There are 298 time series of the error correction terms $\gamma_{i,j,t}$, one for each pair from January $1861$ to December $1869$, with all the posterior medians between $-0.817$ and $-0.003$ for the whole period under our study. The posterior mean of the standard deviation $\sigma_{\gamma}$ is $0.204$ with a $95$\% posterior interval $[0.143, 0.273]$, confirming the need for time-varying coefficients (as time-invariant coefficients would correspond to $\sigma_{\gamma}=0$).

\begin{figure}
    \centering
    \includegraphics{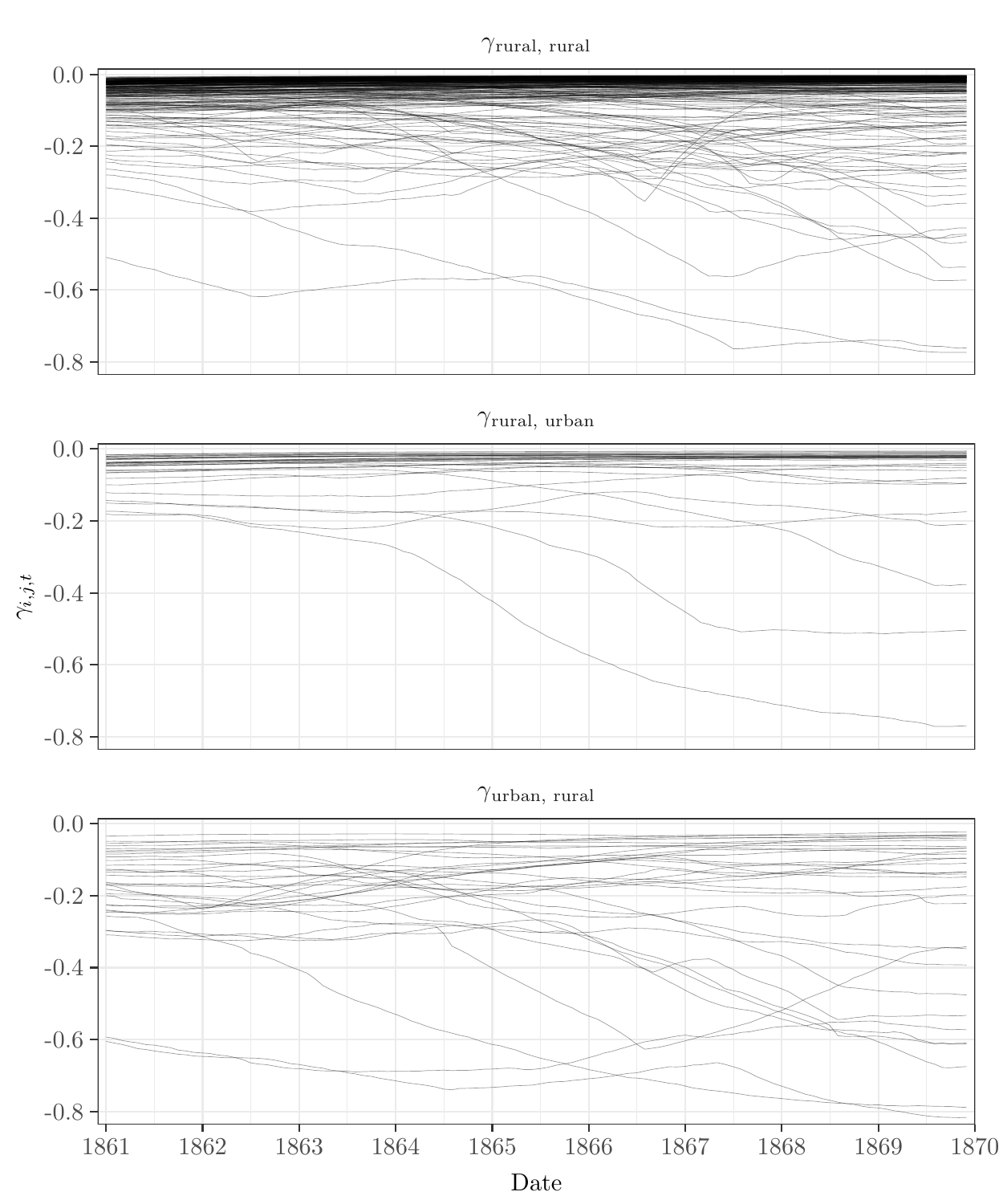}
    \caption{Medians of $12{,}000$ posterior draws for all coefficients $\gamma_{i,j,t}$ by groups.}
    \label{fig:gamma}
\end{figure}

In \autoref{fig:gamma} we present the coefficients $\gamma_{i,j,t}$ using the same grouping as with the coefficients $\beta_{i,j,t}$ in \autoref{fig:beta}. Most of the coefficients $\gamma_{i,j,t}$ are fairly steady over time, and there are no clear thresholds visible marking the start or end of the famine period. The upper panel shows the error correction coefficients for rural--rural pairs and the middle panel for rural--urban pairs. Both reveal that price correction to emerging disequilibrium was slow among the rural markets and between the rural--urban pairs. Furthermore, both show that while there are some pairs with speedier market adjustments during the famine, in the vast majority of rural--rural and rural--urban pairs, there are no worthwhile changes in the price transmission during the time span in question.

\begin{figure}
    \centering
    \includegraphics{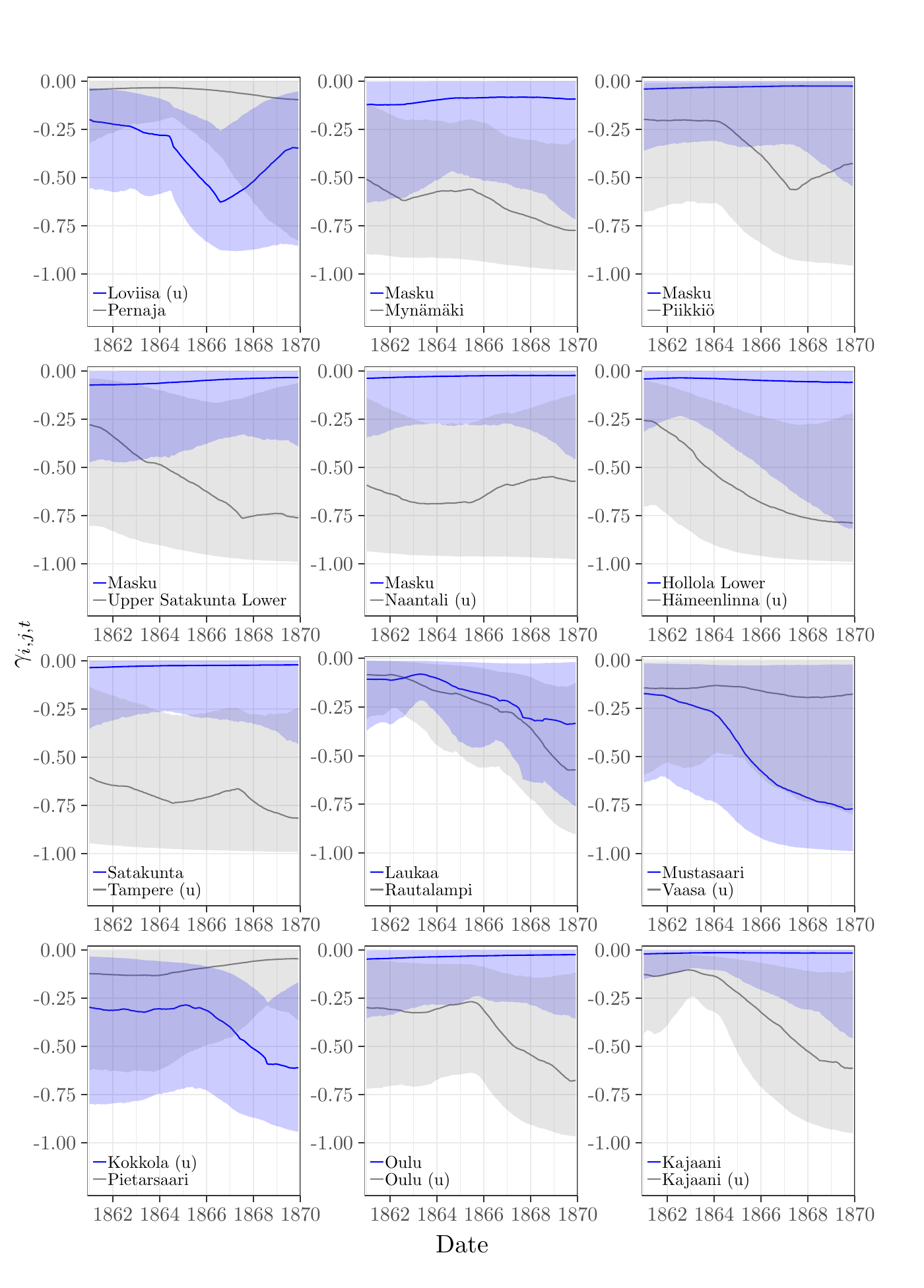}
    \caption{Posterior medians and $95$\% posterior intervals of the coefficients $\gamma_{i,j,t}$ and their counterparts $\gamma_{j,i,t}$ for those pairs where the posterior median is smaller than $-0.55$ at some time point. The name in the legend corresponds to the site $i$ in the coefficient $\gamma_{i,j,t}$.}
    \label{fig:gammagrid}
\end{figure}

The lower panel in \autoref{fig:gamma} shows the coefficients $\gamma_{i,j,t}$ for urban--rural pairs illustrating how the urban prices adjust to the urban--rural price differentials. In many instances, the urban markets adjusted faster to emerging price disequilibrium during the famine than before it. The acceleration of the market adjustment during the famine was most pronounced in northern coastal towns (see supplementary maps on GitHub), but it also occurred in some inland towns. While more marked in the urban markets, the speedier error correction during the famine was not completely confined to towns. There were also some rural regions that witnessed a faster reaction to price differences, many of which were located in south and southwest Finland, with some regions also along the western coast and further inland. Examples of neighbour pairs are displayed in \autoref{fig:gammagrid} where those $\gamma_{i,j,t}$ series with values smaller than $-0.55$ at some time point alongside their counterpart series $\gamma_{j,i,t}$ are plotted.

\autoref{fig:gamma} and \autoref{fig:gammagrid} show that the response of the coefficients $\gamma_{i,j,t}$ to the famine conditions does not happen abruptly, nor do they recover immediately after the famine. This emphasises the benefits of time-varying coefficients. Many of the coefficients $\gamma_{i,j,t}$ begin to decrease in 1865--1866, some even earlier and do not recover before the end of our time span. This means that price transmissions accelerated much before the famine escalating crop failure in September 1867 and continued at that level after the famine had ceded.

\subsection{Visualizing spatial price propagation}\label{sec:maps}

To better understand the overall functioning of the market system and the importance of the close to $600$ pairwise coefficients $\beta_{i,j}$ and $\gamma_{i,j,t}$, we examined the expected values of the latent log-prices $\mu_t$ in response to a price increase in one region. For this purpose, the trend $\alpha_t$ was set to zero for all $t$, and the error correction coefficients were selected from July (the last month uninfluenced by a new harvest) for the years 1861 (before famine) and 1868 (the peak year of the famine). The estimated coefficients $\beta_{i,j}$ were used as they were. The initial log-prices were fixed to equal values (as we were interested in the changes in prices, the actual value could be chosen arbitrarily), except for one specified place where the log-price was increased by $0.01$, corresponding to approximately $1$\% price increase. The expected values were then calculated using the posterior samples of the coefficients $\beta_{i,j}$ and $\gamma_{i,j,t}$ and finally taking an average over all the simulated values of $\mu_{i,t}$. We restricted the simulation to $12$ months. This corresponded to the annual harvest cycle. Furthermore, changes in prices after $12$ months were generally negligible.

\begin{figure}
    \centering
    \includegraphics{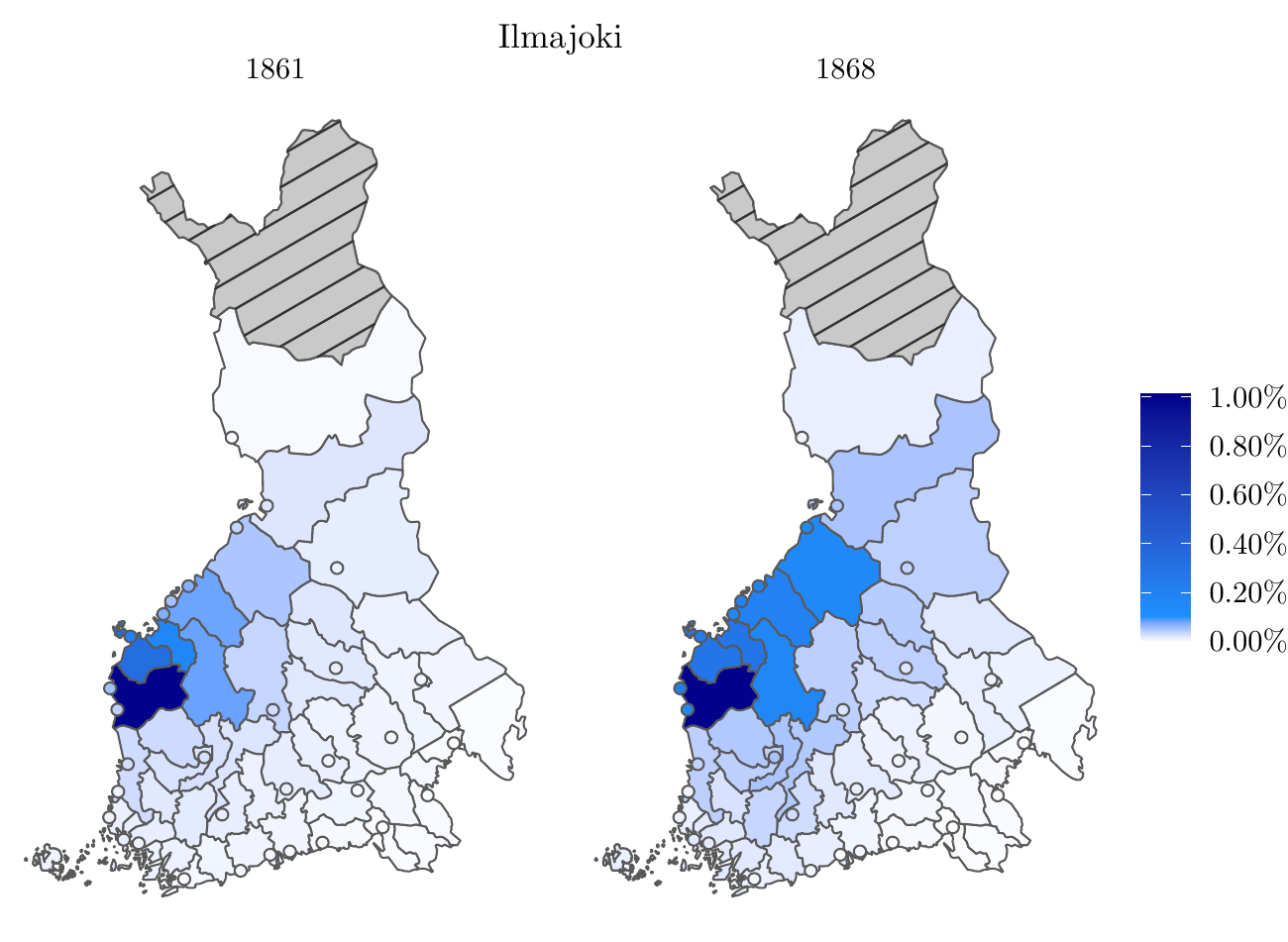}
    \caption{Regional maximum percentage increase in prices during a twelve-month period with respect to the initial price due to a $1$\% increase in price in Ilmajoki (dark blue). Values are based on $\bm{\alpha}=\bf{0}$, posterior samples of $\beta_{i,j}$ and $\gamma_{i,j,t}$, with $t$ fixed to July of the year marked in each panel. The striped grey area is not included in our study.}
    \label{fig:maps_ilmajoki}
\end{figure}

\begin{figure}
    \centering
    \includegraphics{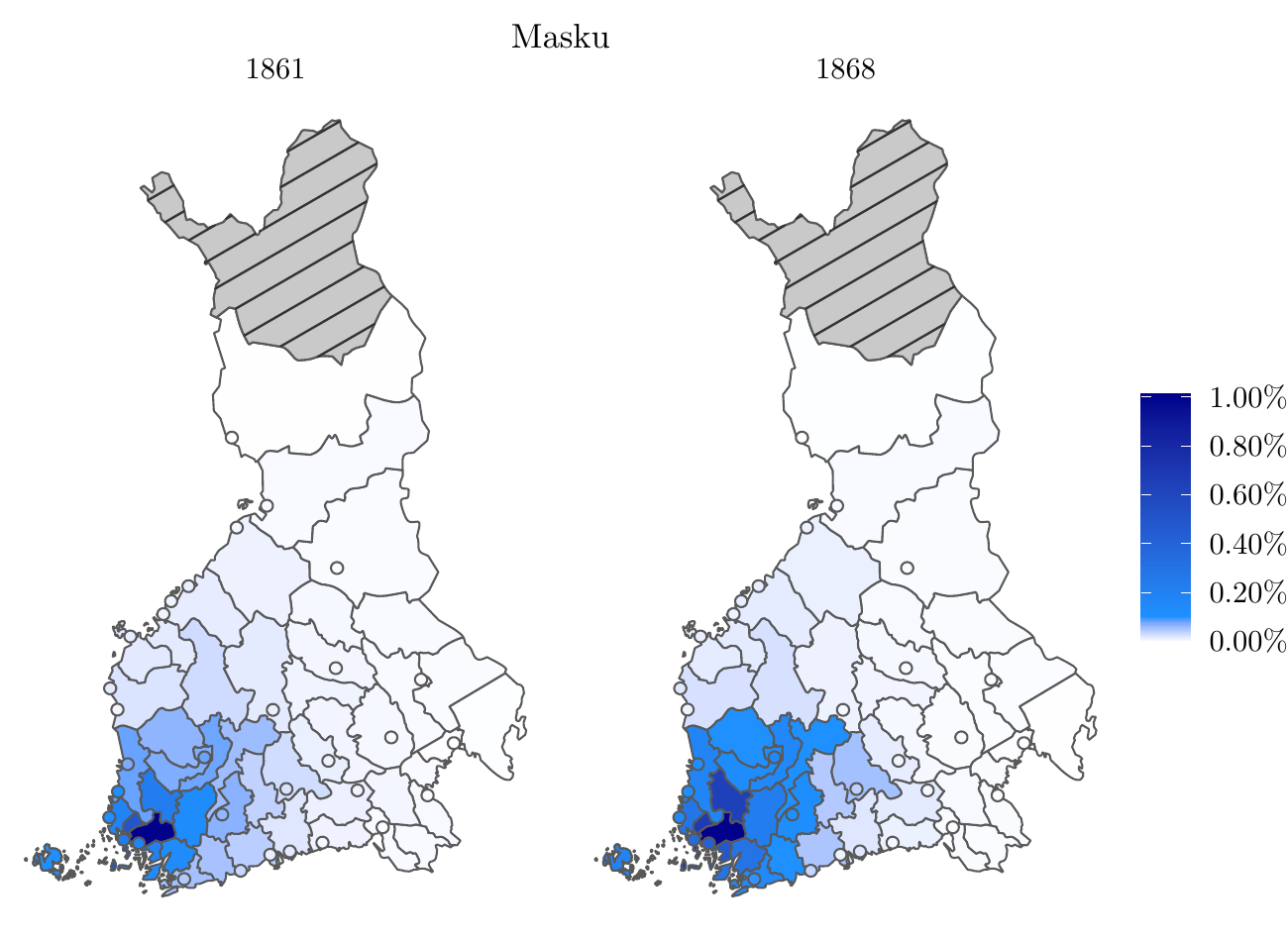}
    \caption{Regional maximum percentage increase in prices during a twelve-month period with respect to the initial price due to a $1$\% increase in price in Masku (dark blue). Values are based on $\bm{\alpha}=\bf{0}$, posterior samples of $\beta_{i,j}$ and $\gamma_{i,j,t}$, with $t$ fixed to July of the year marked in each panel. The striped grey area is not included in our study.}
    \label{fig:maps_masku}
\end{figure}

\begin{figure}
    \centering
    \includegraphics{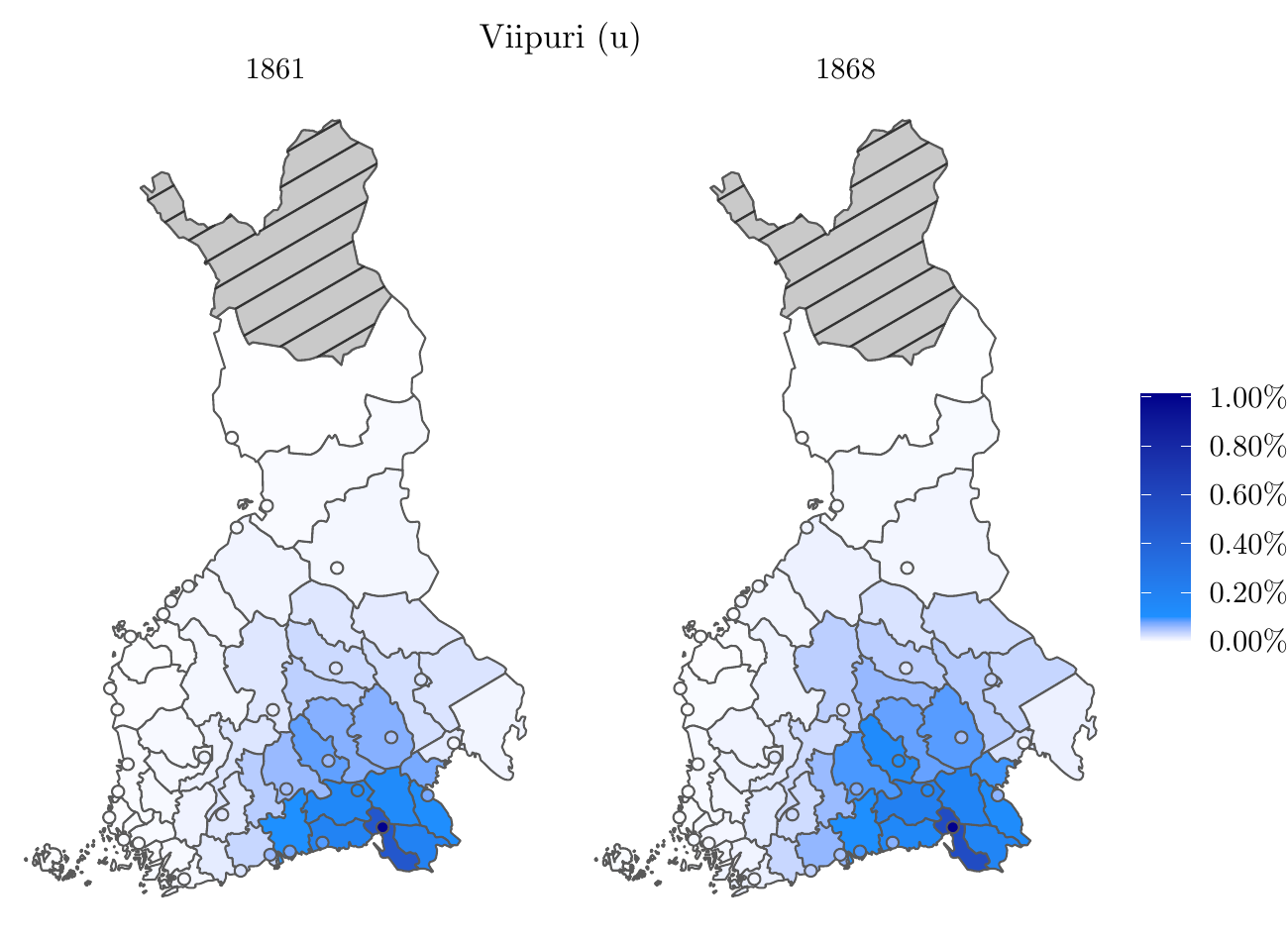}
    \caption{Regional maximum percentage increase in prices during a twelve-month period with respect to the initial price due to a $1$\% increase in price in town of Viipuri (dark blue). Values are based on $\bm{\alpha}=\bf{0}$, posterior samples of $\beta_{i,j}$ and $\gamma_{i,j,t}$, with $t$ fixed to July of the year marked in each panel. The striped grey area is not included in our study.}
    \label{fig:maps_viipuriu}
\end{figure}

Figures \ref{fig:maps_ilmajoki}, \ref{fig:maps_masku}, and \ref{fig:maps_viipuriu} present the maximum percentage increase in regional prices due to a $1$\% increase in price in region $j$. As in \autoref{fig:mu} we used Ilmajoki, Masku, and Viipuri as examples.

\autoref{fig:maps_ilmajoki} depicts the effects of a $1$\% change in the rye price in the administrational district of Ilmajoki in July 1861 and 1868. The resulting maximum price increase is modest, and only in one region it surpasses $0.3$\% with respect to its initial level over the twelve-month window. The price increase is larger north of Ilmajoki, and this tendency strengthens during the famine: in 1868, the price shock travels further northeastward and reaches further inland. The pattern agrees with the increased price transmission that happened in many Ostrobothnian coastal towns during the famine (see \autoref{fig:gamma}, \autoref{fig:gammagrid} and supplementary maps on GitHub). The asymmetry of the spread is distinct: the price increase in the southern neighbouring district of Upper Satakunta Upper is at maximum $0.05$\% (in 1868) in response to a $1$\% shock whereas in the northern neighbouring district of Mustasaari the corresponding value is $0.28$\%. 

\autoref{fig:maps_masku} depicts a price increase in the southwestern district of Masku. Unlike in the case of Ilmajoki, the spatial reach of the shock does not change much during the famine, but the price transmission intensifies close to the shock origin. For example, in the northern neighbouring region of Upper Satakunta Lower, a $1$\% price increase in Masku results in price increases of c. $0.23$\% and $0.64$\% in 1861 and 1868, respectively. The fact that the spatial reach remains constant probably reflects the fact that the southwestern Finnish markets were fairly well integrated to begin with and suggests that well-integrated markets saw little change in their operation during the famine.

\autoref{fig:maps_viipuriu} depicts a shock to the southeastern town of Viipuri. Here, the spread patterns are like those in Ilmajoki. The spatial spread of the shock is asymmetric, with more pronounced westward travel from the shock origin, without apparent difference between the years 1861 and 1868. 

These maps suggest that the price propagation retained their pre-famine routes: the famine typically did not carve out new trade paths. Furthermore, the markets were fairly thin, and price transmission was strong only to places close-by to the shock origin. Further research is needed to tell whether observed route stabilities, spatial asymmetries and lack of long-distance price transmission stemmed from liabilities of existing trade relationships, lack of information, or, for example insurmountable transportation costs. 

\section{Discussion}\label{sec:discussion}

To analyse the grain market integration during the Finnish 1860s famine, we introduced a spatial context to the well-known error correction framework and modeled all the regional price time series simultaneously. This allowed us to omit the complex procedures to predetermine the market leader and the (time-varying) long-term relationships between the series. Also, allowing the common trend and error correction parameters to vary smoothly in time enabled us to estimate the temporal changes in the market integration without relying on artificial, sharp time period demarking the famine. 

Depending on the available data and prior information, our model could be further modified and extended, for example, through more detailed modelling of the trend or error correction terms. For example, the common trend term $\alpha$ could be specified with some functional form of seasonal variation related to, for instance, harvests. We treated the error correction terms as independent random walks, but these could be allowed to depend on each other according to some (spatial) correlation structure or grouping. Time-varying components could also be assumed to vary more smoothly using integrated random walks, or piecewise constant given prior knowledge of potential the change points. It is also possible to let the short-run coefficients vary in time, although this induces a significant increase in the computational burden due to the subsequent time-varying covariance matrix in likelihood computations. Additional data on the regions' characteristics could be incorporated in the model to explain the differences in the short-run and long-run coefficients of the regions for a further insight of the market dynamics. Finally, we have defined the neighbourhood based on border sharing, but it could be interesting to study whether addition of second-order neighbours (neighbours of neighbours) would affect the results, especially for urban towns surrounded by a single rural region. Further extensions to enable simultaneous modelling of multiple products could also be an interesting future venue (an alternative agent-based modelling approach to model multivariate spatio-temporal panel data was used in \citep{Cun2021}).

Regarding the patterns of market integration, we detected certain differences from those in the original contribution of \citet{grada2001markets}. We estimated generally weaker price co-movement and, importantly, documented increased price transmission principally in the urban markets. In addition to modelling differences, it is likely that some of the differences can be attributed to the use of provincial aggregates in \citet{grada2001markets}, which may overemphasise the role of small urban locations.

A possible reason for the increased price transmission during the famine may be the connection between an ill-developed pre-famine market system, the pre-famine prevalence of subsistence farming, and the introduction of deficit producers to the food markets after crop failures \citep{devereux1988entitlements}. The fact that we observed little change in the behaviour of the reasonably well-developed southwestern and southeastern Finnish food markets during the famine aligns with this interpretation.

Our results show that the regional aspects of the early 19th century Finnish grain markets \citep[e.g.,][]{voutilainen2020multi} were still in place during the 1860s famine. The overall weak price transmission not only provides an explanation for the emergence of persistent east-west price gaps observed by \citet{grada2001markets}, but also yields an important wider implication for the study of market integration. The results also suggest that moderate price dispersion (displayed in \autoref{fig:price}) and reasonably high level of sigma convergence (measured in coefficient of variation of regional prices) are not universally coincided with efficient price transmission between the regional markets. Further research is needed to understand how low levels of price dispersion were achieved in this kind of setting and whether it was driven by certain key markets or reasonable symmetry of harvest outcomes. Our results show that the trade routes appeared robust to a large-scale harvest shock of 1867. Markets rarely changed their spatial orientation in response to this, even though the harvest failures came with substantial geographic variation. Transportation costs, lack of information, and preexisting trade connections probably explain this. 

Where markets behaved better during the famine, they facilitated the spread of the price shocks. The thin markets had the advantage of limiting the shocks to locations close by. The increased market transmission during the famine proliferated the spread of the shocks, especially along the western coast.

\section*{Acknowledgments}
Tiia-Maria Pasanen was supported by the Finnish Cultural Foundation. Jouni Helske was supported by the Academy of Finland grants 331817 and 311877. Miikka Voutilainen was supported by the Academy of Finland grant 308975. The authors wish to acknowledge CSC – IT Center for Science, Finland, for computational resources.

\bibliographystyle{abbrvnat} 

\bibliography{arxiv}

\end{document}